\documentclass[preprint,showpacs,aps,prd,nofootinbib,letterpaper]{revtex4-1}
\usepackage{graphicx}
\usepackage{subfigure}
\usepackage{dcolumn}
\usepackage{amsmath}
\usepackage{amssymb}
\usepackage{times}
\usepackage{epsfig}
\usepackage{float}     
\usepackage{relsize}
\usepackage{mathrsfs}
\RequirePackage{xspace}

\begin{document}
\title{The coherent cross section of vector mesons in ultraperipheral PbPb collisions at the LHC}
\author{Ya-ping Xie}\email{xieyaping@impcas.ac.cn}
\affiliation{Institute of Modern Physics, Chinese Academy of
Sciences, Lanzhou 730000, China}
\affiliation{Department of Physics, Lanzhou University, Lanzhou 730000, Chnia}
\affiliation{Key Laboratory of Quark and Lepton Physics (MOE) and Institute
 of Particle Physics,
Central China Normal University, Wuhan 430079, China
}
\author{Xurong Chen}\email{xchen@impcas.ac.cn}
\affiliation{Institute of Modern Physics, Chinese Academy of
Sciences, Lanzhou 730000, China}
\begin{abstract}
The coherent cross section of $J/\psi$, $\rho$, $\phi$ are computed in dipole model in the ultraperipheral PbPb collisions, the IP-Sat and IIM model are applied in the calculation of the differential cross section of the dipole scattering off the nucleon, three kinds of forward vector meson wave functions are used in the overlap. The prediction of  $J/\psi$ and $\rho$  is compared with the experimental data of the ALICE Collaboration, and the prediction of $\phi$ is also given in this paper.
\end{abstract}
\maketitle
\section{Introduction}
 The production of vector mesons through a virtual photon-hadron scattering had been studied in Deep Inelastic Scattering at HERA~\cite{Ivanov:2004ax}, the ultraperipheral collisions at the LHC offer an interesting way to study the  photonproduction of vector mesons through a real photon scattering off a hadron at the high energy limit~\cite{Bertulani:2005ru,Baltz:2007kq},  recently, the ALICE Collaboration have measured vector mesons production in PbPb ultraperipheral collisions~\cite{Aaij:2013jxj,Abbas:2013oua,TheALICE:2014dwa,Adam:2015gsa,Abelev:2012ba}. On the theoretical front, the photon production of vector mesons have been studied through various approaches, including perturbative quantum chromodynamics(pQCD), $k_T$-factorization and color dipole model(CDM)
~\cite{Klein:1999qj,Frankfurt:2002sv,Goncalves:2005yr,Rebyakova:2011vf,Adeluyi:2012ph,Cisek:2012yt,Ducati:2013bya,Lappi:2013am,Guzey:2013qza}, in this work, we use the dipole model to predict the vector mesons production in PbPb ultraperipheral collisions at $\sqrt{s_{NN}}$=2.76TeV, the goal of this work is to update the prediction of vector mesons with new fit of IP-Sat model and several wave function models.\\
\indent  According to the dipole model, the process of the photon-hadron scattering can be viewed as three steps, at the first step, the virtual or real photon splits into a  dipole with quark and antiquark, at the second step, the dipole scatters off the hadrons, at the last step, the dipole becomes a vector meson. The amplitude of the photon-hadrons scattering contains three portions, the light-cone wave function of the photon splitting into dipole, the cross section of the dipole scattering off the proton, and the forward wave function for vector mesons. The  calculations of cross section of the dipole scattering off a proton is relative to the gluon distribution in the small-x region, in the literature, various parameterization models have successfully implement to calculate the cross section of the dipole scatters off a proton, such GBW model ~\cite{GolecBiernat:1998js,GolecBiernat:1999qd}, IP-Sat model~\cite{Bartels:2002cj,Kowalski:2003hm,Kowalski:2006hc,Rezaeian:2012ji} and IIM model~\cite{Iancu:2003ge,Soyez:2007kg,Watt:2007nr,Rezaeian:2013tka}. The light-cone wave function of the photon splitting into the dipole can be calculated in QED, but the forward wave function of the vector meson can not be calculated analytically, the forward wave functions of vector meson almost are modeled the light-cone wave function of the photon, the models of the vector mesons include the Gaus-LC~\cite{Kowalski:2006hc}, DGKP~\cite{Dosch:1996ss,Goncalves:2004bp}, Boosted Gaussian~\cite{Nemchik:1996cw,Forshaw:2003ki} and so on.\\
 \indent This paper is organized as follow,  the brief review of the dipole model and wave function will be presented in Sec II, the numerical results and some discussion will be presented in Sec III.
\section{The coherent vector meson cross section}
\subsection{Formulas in the ultraperipheral collision }
This work, we consider the coherent cross section of the vector meson in the PbPb ultraperipheral collisions, in hadronic collisions, when the impact parameter is larger, the two hadrons almost don't touch each other, but the real photon can be emitted from the hadrons in the high energy limit, therefore, the real photon can scatter off the hadrons, in this process, the rapidity distribution  can be factorized into the equivalent photon flux and the cross section of the photon-hadrons scattering, the formula is 
\begin{eqnarray}
\frac{d\sigma^{h_1h_2}}{dy}=\bigg[n^{h_1}(\omega)\sigma^{\gamma h_2}(\omega)\bigg]_{\omega_{left}}+\bigg[n^{h_2}(\omega)\sigma^{\gamma h_1}(\omega)\bigg]_{\omega_{right}},
\label{dndk}
\end{eqnarray}
where  $y $  is the rapidity of the vector meson, the  $\sigma^{\gamma A}(\omega)$ is the cross section of the photon-hadrons scattering, the $n(\omega)$ is the equivalent photon flux in the hadrons,  with $\omega_{left}=\frac{M_v}{2}\exp(-y)$, and $\omega_{right}=\frac{M_v}{2}\exp(y)$, whtere $M_V$ is the mass the vector meson. In proton-proton scattering, the equivalent photon flux is~\cite{Drees:1988pp}
   \begin{eqnarray}
   n(\omega)=\frac{\alpha_{em}}{2\pi}\bigg[1+\bigg(1-\frac{2\omega}{
   \sqrt{s_{NN}}}\bigg) \bigg]\bigg[\ln D-\frac{11}{6}+\frac{3}{D}-\frac{3}{2D^2}+\frac{1}{3D^3} \bigg],
   \end{eqnarray}
   where $\sqrt{s_{NN}}$ is the nucleon-nucleon center energy,  $D=1+\frac{0.71\mathrm{GeV}^2}{Q^2_{min}}$, with $Q^2_{min}=\omega^2/\gamma_L^2$, $\gamma_L$ is the lorentz factor, with $\gamma_L=\sqrt{s_{NN}}/2m_p$. In the nucleus-nucleus scattering, the equivalent photon flux is~\cite{Klein:1999qj}
\begin{eqnarray}
n(\omega)=\frac{2 Z^2\alpha_{em}}{\pi}\big[\xi K_1(\xi)
K_0(\xi)-\frac{\xi^2}{2}[K_1^2(\xi)-K_0^2(\xi)]\big],
\end{eqnarray}
 where   $\xi=2\omega R_A/\gamma_L$, with $R_A$ is the radius of the nucleus, $K_0(x)$ and $K_1(x) $ are the second kind of Bessel functions. \\
 \indent  The $\sigma(\omega)$ is the cross section of the photon-hadrons scattering, it can be integrated from the differential cross section, the coherent differential cross section of the photon-proton scattering is calculated as~\cite{Kowalski:2003hm,Kowalski:2006hc} 
\begin{eqnarray}
\frac{d\sigma^{\gamma p\to Vp}}{dt}=\frac{R_g^2(1+\beta^2)}{16\pi}
\left|\mathcal{A}^{\gamma p\to Vp}(x_p, Q^2,\Delta)\right|^2,
\label{dsigma1}
\end{eqnarray}
where the amplitude is computed as
\begin{eqnarray}
\mathcal{A}^{\gamma p\to Vp}(x_A, Q^2,\Delta)= i\int d^2r\int_0^1\frac{dz}{4\pi}
\int d^2b(\Psi_V^*\Psi_{\gamma})_{T}(z,r,Q^2)e^{-i(b-(1-z)r)\cdot \Delta }\frac{d\sigma_{q\bar{q}}}{d^2b}.
\end{eqnarray}
Where $t=-\Delta^2$, the relationship between the $x_p$ amd $y$ is $x_p=M_v\exp(-y)/\sqrt{s_{NN}}$ and  $T$denotes the transverse overlap of the wave functions of the photon and vector meson, the $\mathcal{N}(x,r,b)$ is the amplitude of the dipole scattering off the nucleon, which will be considered in the next subsection. The factor
$\beta $ is the ratio of the real part to the imaginary part of amplitude, it is computed as
\begin{equation}
\beta=\tan(\frac{\pi}{2}\delta),
\end{equation}
where $\delta$ is calculated as
\begin{equation}
\delta=\frac{\partial \ln (\mathrm{Im}\mathcal{A}(x))}{\partial \ln1/x}.
\end{equation}
The factor $R_g^2$ reflects  the skewdness, it gives~\cite{Shuvaev:1999ce} 
\begin{equation}
R_g=\frac{2^{2\delta+3}}{\sqrt{\pi}}\frac{\Gamma(\delta+5/2)}{\Gamma(\delta+4)}.
\end{equation}
\indent The differential cross section of $\gamma A\to VA$ is written as
\begin{eqnarray}
\frac{d\sigma^{\gamma A\to VA}}{dt}=\frac{R_g^2(1+\beta^2)}{16\pi}
\left|\langle\mathcal{A}^{\gamma A\to VA}(x_p, Q^2,\Delta)\rangle_N\right|^2,\label{dsigma}
\end{eqnarray}
where the average amplitude is calculated as~\cite{Kowalski:2008sa,Caldwell:2010zza,Lappi:2010dd}
\begin{eqnarray}
\langle\mathcal{A}^{\gamma A\to VA}(x_p, Q^2,\Delta)\rangle_N&=& i\int d^2r\int_0^1\frac{dz}{4\pi}
\int d^2b(\Psi_V^*\Psi_{\gamma})_{T}(z,r,Q^2)e^{-i(b-(1-z)r)\cdot \Delta }\notag\\
&&\times2(1-\exp(-2\pi B_pAT_A(b)\mathcal{N}(x_p,r)).
\label{amp}
\end{eqnarray}
The shape function is defined as
\begin{eqnarray}
T_A(b)=\int_{-\infty}^{\infty}dz\rho_A(\sqrt{b^2+z^2}),
\end{eqnarray}
with Wood-Saxon distribution
\begin{equation}
\rho_A(r)=\frac{N}{\exp(\frac{r-R_A}{\delta_0})+1},
\end{equation}
where $\delta_0=0.54$fm, $R_A=(1,12\mathrm{fm})A^{1/3}-(0,86
\mathrm{fm})A^{-1/3}$, A is the number of nucleus. 
\subsection{The IP-Sat and IIM model}
There are various approaches to calculate the cross section of dipole scattering off the proton,
GBW model was proposed by Golec-Biernat and W\"usthoff~\cite{GolecBiernat:1998js,GolecBiernat:1999qd}, but the GBW model has a shortcoming that it do not match DGLAP evolution equation at large $Q^2$. Then, the Impact Parameter saturation (IP-Sat) model was proposed according to the DGLAP evolution equation, the amplitude of the IP-Sat model reads~\cite{Kowalski:2008sa,Caldwell:2010zza,Lappi:2010dd}
 \begin{equation}
 \frac{d\sigma_{q\bar{q}}}{d^2b}=2[1-\exp(-\frac{1}{2\pi B_p}\frac{\pi^2}{2N_c}r^2\alpha_s(\mu^2)xg(x, \mu^2)T_p(b)],
 \end{equation}
 where  the $T_p(b)$ is defined
  \begin{eqnarray}
  T_p(b)=\exp(-\frac{b^2}{2B_p}).
  \end{eqnarray}
  The scale $\mu^2$ has relationship with the dipole size $r$, 
\begin{equation}
\mu^2=\mu^2_0+\frac{C}{r^2},
\end{equation}
with $C=4$, where the $xg(x,\mu^2)$ is the gluon distribution in the proton, the initial gluon distribution is 
\begin{equation}
xg(x, \mu^2_0)=A_gx^{-\lambda_g}(1-x)^{5.6}.
\end{equation}
The parameters $A_g$, $\mu^2_0$, $\lambda_g$, $B_p$ are determined from the fit to experimental data 
$F_2$, we take the values of parameters according to Ref.~\cite{Rezaeian:2012ji}, There are two sets of the parameters, which are  different from the Ref.~\cite{Kowalski:2003hm,Kowalski:2006hc}, especially for the mass of the light quarks, in the fit of Kowalski et al, the mass of the light quark is $m_q$=0.14~GeV, in the fit of Rezaeian et al, the mass of the light quarks is $m_q\approx$0 GeV.\\
 \begin{table}[H]
 \begin{tabular}{p{2cm}p{2cm}p{2cm}p{2cm}p{2cm}p{2cm}p{2cm}}
 \hline
 \hline
 &$B_p$&$m_{u,d,s}$& $m_c$ & $\mu_0^2$ & $A_g$ & $\lambda_g$\\
 \hline
Para 1 &$4.0$~$\mathrm{GeV}^2$ &$\approx 0  $~GeV& 1.27~GeV     &1.51~GeV$^2$&2.308 &  0.058      \\
 Para 2&$4.0$~$\mathrm{GeV}^2$    &$\approx 0$~GeV    &1.4~GeV		 & 1.428~GeV$^2$    & 2.373&     0.052     \\
  \hline
 \hline
 \end{tabular}
 \caption{The parameters of IP-Sat model~\cite{Rezaeian:2012ji}}
 \label{IPP}
 \end{table} 
\indent In Ref.~\cite{Bartels:2002cj,Lappi:2010dd}, the authors used a factorized impact parameter saturation model, it reads
  \begin{eqnarray}
\frac{d\sigma_{q\bar{q}}}{d^2b}& \approx& 2T_p(b)\mathcal{N}(x,r)\notag\\
  &=&2T_p(b)[1-\exp(-\frac{1}{2\pi B_p}\frac{\pi^2}{2N_c}r^2\alpha_s(\mu^2)xg(x, \mu^2)].
  \end{eqnarray}
  we use fIP-Sat model to calculate the vector meson cross section in photon-nucleus scattering.\\
\indent On the other side, Iancu, Itakura and Munier proposed a saturation model based on the solution to BK evolution equation~\cite{Iancu:2003ge}, we use the impact parameter dependent saturation model, we take the form as the same as Refs.~\cite{Soyez:2007kg,Lappi:2010dd}
  \begin{eqnarray}
\frac{d\sigma_{q\bar{q}}}{d^2b}=2T_b(b)\mathcal{N}(x,r),
  \end{eqnarray}
 the amplitude is written as 
 \begin{eqnarray}
 \mathcal{N}(x,r)=\begin{cases}
 \mathcal{N}_0(\frac{rQ_{s}}{2})^{2(\gamma_s+(1/\kappa\lambda Y)\ln(2/rQ_s))},\quad rQ_s\le2,\\
 1-\exp\big(-a\ln^2(b rQs)\big),\quad\quad\quad\!\!\! rQ_s>2.
 \end{cases}
 \end{eqnarray}
 with $Y=\ln(1/x)$ and $\kappa=9.9$, where $Qs(x,b)=(x_0/x)^{\lambda/2}$~GeV,  $a$ and $b$ are
 \begin{equation}
 \begin{split}
& a=-\frac{\mathcal{N}^2_0\gamma_s^2}{(1-\mathcal{N}_0)^2\ln(1-\mathcal{N}_0)},\\
 &b=\frac{1}{2}(1-\mathcal{N}_0)^{-(1-\mathcal{N}_0)/(2\mathcal{N}_0\gamma_s)}.
 \end{split}
 \end{equation}
 The parameters $ B_p, \mathcal{N}_0, \gamma_c, \lambda, x_0$ are need to be determined from the fit to the experimental data $F_2$, we take the parameter as the same as Ref.~\cite{Soyez:2007kg}, they are presented in the following Table:
 \begin{table}[h]
 \begin{tabular}{p{2cm}p{2cm}p{2cm}p{1.5cm}p{1.5cm}p{1.5cm}p{1.5cm}p{3cm}}
 \hline
 \hline
 $B_p$&$m_{u,d,s}$& $m_c$ & $m_b$& $\mathcal{N}_0$ & $\gamma_c$ & $\lambda$& $x_0$\\
 \hline
 $5.59$~GeV$^{-2}$ &$0.14$~GeV& 1.4~GeV     &4.5~GeV & 0.7 &0.7376 &0.2197&$ 1.632\times10^{-4} $\\
 \hline
 \hline
 \end{tabular}
 \caption{The parameters of the IIM model~\cite{Soyez:2007kg}.}
 \label{IIMP}
 \end{table} 
 
 \subsection{The forward vector meson wave functions}
\indent The $(\Psi_V^*\Psi_{\gamma})_{T}(r,z)$  is the transverse overlap of the functions of vector meson and the photon, there are various models for the forward vector meson wave function in the literature, in this work, we take the three kinds of model for the vector meson wave functions.  At first, we consider the Boosted Gaussian and Gaus-LC model, the overlap takes following form in Boosted Gaussian and Gaus-LC model~\cite{Kowalski:2006hc},
\begin{eqnarray}
(\Psi_V^*\Psi_{\gamma})_T(r,z)=e_fe\frac{N_c}{\pi z(1-z)}\lbrace  m_f^2
K_0(\epsilon r)\phi_T(r,z)-(z^2+(1-z)^2)\epsilon K_1(\epsilon r)\partial_r
\phi_T(r,z)\rbrace ,\notag\\
\end{eqnarray}
where $e=\sqrt{4\pi\alpha_{em}}$,  $m_f$ is the mass of quarks, $e_f$ is the electric charge of the  quarks, $\epsilon=\sqrt{z(1-z)Q^2+m_f^2}$, $N_c$ is the number of the colors. The scalar function  $\phi_T(r,z)$  of Gaus-LC model~\cite{Kowalski:2006hc} reads 
\begin{eqnarray}
\phi_T(r,z)=N_T(z(1-z))^2\exp(-\frac{r^2}{2R_T^2}),
\end{eqnarray}
The Boosted Gaussian model is simplified from NNPZ model~\cite{Nemchik:1996cw,Forshaw:2003ki}, the scalar function of Boosted Gaussian reads
\begin{eqnarray}
\phi_T(z,r)=N_Tz(1-z)\exp\big(-\frac{m_f^2\mathcal{R}^2}{8z(1-z)}-
\frac{2z(1-z)r^2}{\mathcal{R}^2}+\frac{m_f^2\mathcal{R}^2}{2}\big).
\end{eqnarray}
DGKP model is another famous model for the forward vector meson wave function~\cite{Dosch:1996ss,Goncalves:2004bp,Forshaw:2003ki}, in this work, we also consider the contribution of DGKP model, the overlap is different from the above models, in  the DGKP model, the overlap reads~\cite{Forshaw:2003ki}
\begin{eqnarray}
(\Psi_V^*\Psi_{\gamma})_T(r,z)=\frac{e f_v}{M_V}f_T(z)\exp(-\frac{\omega_T^2r^2}{2})
\{(\omega_T^2\epsilon r[(z^2+(1-z)^2)]K_1(\epsilon r)+m_f^2K_0(\epsilon r)\},\notag\\
\end{eqnarray}
where $f_v$ is the decay constant of the vector meson, the $f(z)$ reads
\begin{equation}
f_T(z)=N_T\sqrt{z(1-z)}\exp(-\frac{M_V^2(z-1/2)^2}{2\omega^2_T}).
\end{equation}
The parameters of the vector meson functions are determined by the normalization condition and the decay constant, we present the parameters in Table.~\ref{WP}, some parameters are taken from the Ref.~\cite{Kowalski:2006hc, Rezaeian:2013tka}.
\begin{table}[h]
\begin{tabular}{p{1cm} | p{1.5cm}p{1.5cm}p{1.5cm}p{1.5cm} | p{1.5cm}p{1.5cm}  | p{1.5cm}p{1.5cm}|p{1.5cm}p{1.5cm}}
\hline
\hline
meson &$e_f$& mass& $f_v$ & $m_f$& $N_T$& $R_T^2$& $N_T$& $\mathcal{R}^2$& $N_T$& $\omega_T$\\
\hline
&&GeV &GeV & GeV &   &GeV$^2$ & &GeV$^2$ & & GeV\\
\hline
$J/\psi$    &$2/3$    & 3.097    & 0.274  & 1.4      & 1.23       & 6.5 & 0.578      & 2.3  & 8.264     & 0.56\\
$J/\psi$  &$2/3$      & 3.097    & 0.274  & 1.27      & 1.45       & 5.57 & 0.60      & 2.36& 9.18      & 0.568\ \\
$\phi$     &$1/3$     & 1.019    & 0.076  &  0.14    & 4.75       & 21.9& 0.919       & 11.2  &12.12      & 0.269\\
$\phi$     &$1/3$     & 1.019    & 0.076  &  0.01    & 5.91       & 16.45& 1.021       & 11.4  &14.81     & 0.268\\
  $\rho$       &$1/\sqrt{2}$  &   0.776 & 0.156 & 0.14 	  & 4.47	 	& 21.9  & 0.911	 	& 12.9& 8.62	 	& 0.223\\
  $\rho$       &$1/\sqrt{2}$  &   0.776 & 0.156 & 0.01 	  & 5.89	 	& 21.68  & 1.004	 	& 13.3& 11.27	 	& 0.222\\
\hline
\hline
\end{tabular}
\caption{The parameter of the wave functions, the column 6, 7 are the parameters of Gaus-LC, the column 8,9 are parameters of Boosted Gaussian model, the column 10,11 are parameters of DGKP model. }
\label{WP}
\end{table} 

\section{results and discussion }
\begin{figure}[H]
\begin{center}
\includegraphics[width=9cm]{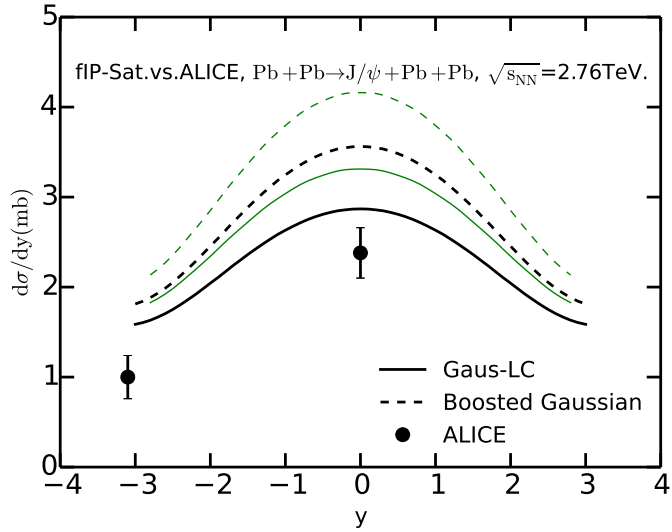}\hspace{3.0cm}
\caption{(Color online) The coherent $J/\psi$ rapidity distribution in PbPb collision at $\sqrt{s_{NN}}$=2.76TeV computed using fIP-Sat model and compared to the experimental data of ALICE~\cite{Abelev:2012ba,Abbas:2013oua}, the black thick lines are using the parameters of fIP-Sat model parameter set 2 in Table.~\ref{IPP}, the green thin curve are the results from Ref.~\cite{Lappi:2013am} .}\label{fig1}
\end{center}
\end{figure}

\begin{figure}[H]
\begin{center}
\includegraphics[width=5in]{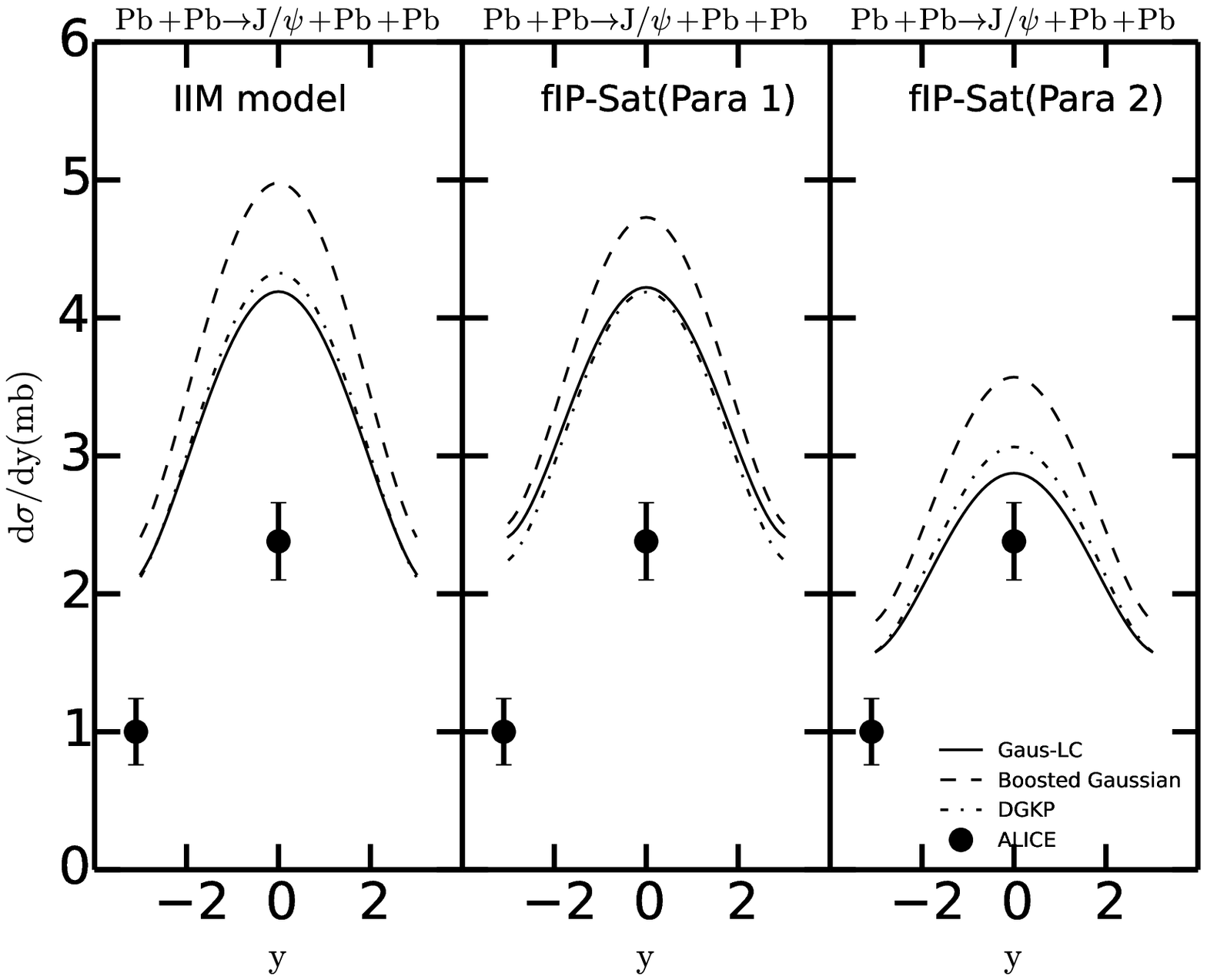}\hspace{3.0cm}
\caption{The coherent $J/\psi$ rapidity distribution in PbPb collision at $\sqrt{s_{NN}}$=2.76TeV  computed using IIM and fIP-Sat model with Gaus-LC (solid), Boosted Gaussian(dashed), DGKP(dot-dashed) and compared to  the experimental data of ALICE~\cite{Abelev:2012ba,Abbas:2013oua}.}\label{fig2}
\end{center}
\end{figure}

\begin{figure}[H]
\begin{center}
\includegraphics[width=5in]{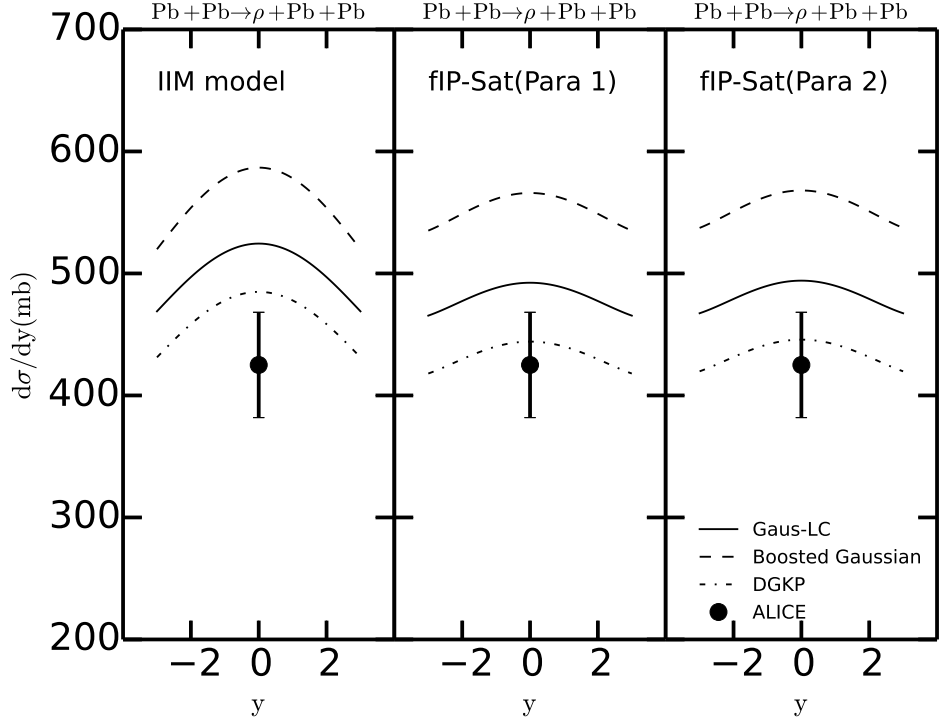}\hspace{3.0cm}
\caption{The coherent $\rho$ meson rapidity distribution in PbPb collision at $\sqrt{s_{NN}}$=2.76~TeV  computed using IIM and fIP-Sat model with Gaus-LC (solid), Boosted Gaussian(dashed), DGKP(dot-dashed) and compared to  the experimental data of ALICE~\cite{Aaij:2013jxj}.}\label{fig3}
\end{center}
\end{figure}

In this section, we shall give our prediction using the fIP-Sat and IIM model with different kinds of wave functions, and compare the prediction to the experimental data. In the calculation using IIM model, we take the mass of charm quark as $m_c=$1.4~GeV, and the mass of the light quarks as $m_q$=0.14~GeV.  In the calculation using fIP-Sat model, we take the mass of quarks as two parameters sets, in parameter set 1, the quark mass is $m_c = $ 1.27~GeV and $m_q=$0.01~GeV, in parameter set 2, the  quark mass is $m_c = $1.4~GeV and $m_q=$0.01~GeV~\cite{Armesto:2014sma}.  The parameters of the wave functions are taken according to the quark mass in fIP-Sat or IIM model, which are presented in Table.~\ref{WP}, the $Q^2= $0~$\mathrm{GeV}^2$ in all calculation in this work because the photon is real photon.\\
\indent In Fig.~\ref{fig1}, we present the prediction of rapidity distribution of $J/\psi$ in PbPb at $\sqrt{s_{NN}}= $2.76~TeV, the prediction was also calculated using the parameters in Ref.~\cite{Kowalski:2006hc}. We can see that the result of this work is lower than the result of Ref.~\cite{Lappi:2013am}, because the newer fit of IP-Sat model in Ref.~\cite{Rezaeian:2012ji} is more accurate than the older fit in the Ref.~\cite{Kowalski:2006hc}.
The prediction using IIM model is also calculated in this work, which is compared with two sets parameters of fIP-Sat model, the results are presented in Fig.~\ref{fig2}, we can see that result of parameter set 2 is closer to the experimental data than the result of IIM model and parameter set 1, the result of Gaus-LC wave function is closer than the Boosted Gaussian and DGKP model.\\ 
\indent The rapidity distribution of $\rho$ meson was also measured at ALICE\cite{Aaij:2013jxj},  the prediction had been presented in Ref.~\cite{Frankfurt:2015cwa},  we also compute the prediction of $\rho$ meson, which is showed in Fig.~\ref{fig3}, we compute the result of $\rho$ meson using IIM model and fIP-Sat with two sets of parameters, we take the light quarks mass $m_q$= 0.14~GeV in the IIM model and take the light quark mass $m_q$= 0.01~GeV in fIP-Sat model. 
We also give the prediction of the rapidity distribution of $\phi$ meson in PbPb collision at $\sqrt{s_{NN}}=$2.76~TeV, there is no experimental data for the $\phi$ meson, we expect the  experimental data of the $\phi $ meson at the LHC in the furture.
\begin{figure}[H]
\begin{center}
\includegraphics[width=5in]{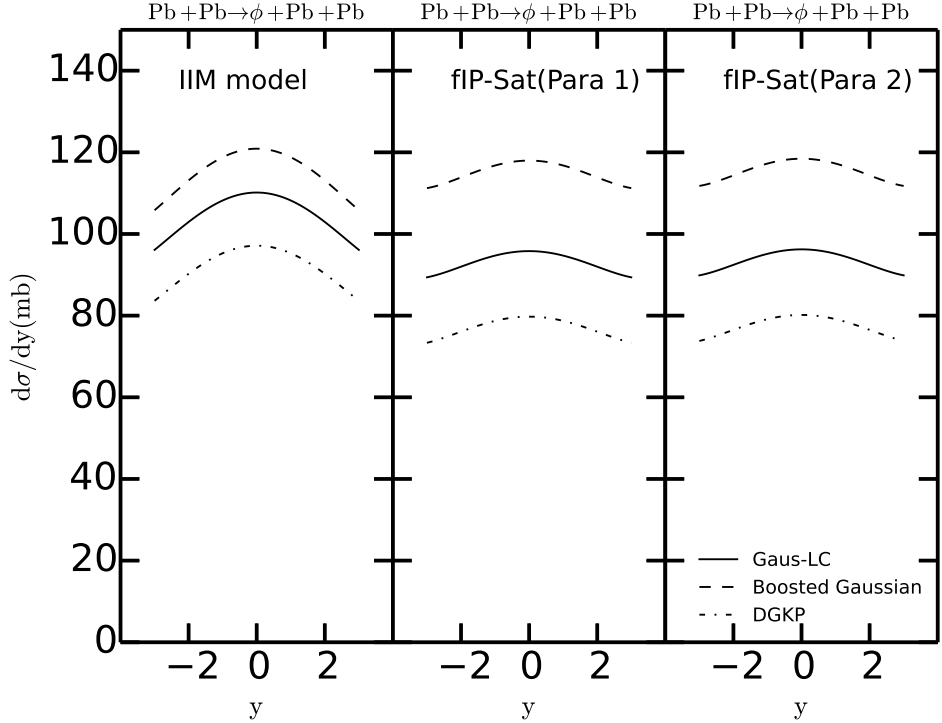}\hspace{3.0cm}
\caption{The coherent $\phi$ rapidity distribution in PbPb collision at $\sqrt{s_{NN}}=$2.76~TeV  computed using IIM and fIP-Sat model with Gaus-LC (solid), Boosted Gaussian(dashed), DGKP(dot-dashed).}.\label{fig4}
\end{center}
\end{figure}
\indent Summary, we calculate the coherent cross section of vector mesons in PbPb ultraperipheral collisions
with fIP-Sat and IIM model, the parameters of this work are determined in fit of HERA data.  We can find that the newer fit of IP-Sat model is more accurate than the older fit
in the calculation of $J/\psi$  production. The IIM model is little upper than the fIP-Sat model in $J/\psi$ and $\rho$ calculations, the production of $\phi$ is also calculated in this work and we hope in the future the production will be measured at the LHC. 
\section{Acknowledgements}
One of the authors, Y. P. Xie, thanks  
 communication with Bo-Wen ~Xiao, T.~Lappi, H.~Mantysaari and A. H. ~Rezaenian. This work is supported in part by the National Natural Science Foundation
of China (Grant No. 11175220), the One Hundred Person
Project (Grant No. Y101020BR0), and the Key Laboratory
of Quark and Lepton Physics (MOE), Central China
Normal University (Grant No. QLPL201414).

\end{document}